\documentclass[twocolumn,showpacs,amsfonts,amssymb,amsmath,floats,superscriptaddress,aps]{revtex4-1}

\usepackage{graphicx}
\usepackage{upgreek}

\usepackage{color}
\usepackage{xfrac}
\usepackage{siunitx}

\def\bra#1{\left.\left\langle#1\right.\right|}
\def\ket#1{\left.\left|#1\right.\right\rangle}

\newcommand{\mr}{\mathrm}

\begin{document}

\title{Signatures of long-range spin-spin interactions in an (In,Ga)As quantum dot ensemble}

\author{Andreas Fischer}
\address{Theoretische Physik II, Technische Universit\"at Dortmund, 44221 Dortmund, Germany}

\author{Eiko Evers}
\address{Experimentelle Physik II, Technische Universit\"at Dortmund, 44221 Dortmund, Germany}

\author{Steffen Varwig}
\address{Experimentelle Physik II, Technische Universit\"at Dortmund, 44221 Dortmund, Germany}

\author{Alex Greilich}
\address{Experimentelle Physik II, Technische Universit\"at Dortmund, 44221 Dortmund, Germany}

\author{Manfred Bayer}
\address{Experimentelle Physik II, Technische Universit\"at Dortmund, 44221 Dortmund, Germany}
\address{Ioffe Physical-Technical Institute, Russian Academy of Sciences, 194021 St. Petersburg, Russia}

\author{Frithjof B.\ Anders}
\address{Theoretische Physik II, Technische Universit\"at Dortmund, 44221 Dortmund, Germany}

\date{\today}

\begin{abstract}

We present an investigation of the electron spin dynamics in an ensemble of singly-charged quantum dots
subject to an external magnetic field and laser pumping with
circularly polarized light.
The spectral laser width is tailored such that different groups of quantum dots are
coherently pumped.
Surprisingly, the dephasing time $T^*$ of the electron spin polarization depends
only weakly on the laser spectral width. 
These findings can be consistently explained by a cluster theory of coupled quantum
dots with a long range electronic spin-spin interaction.
We present a numerical simulation of the spin dynamics based on the central spin
model that includes a quantum mechanical description of the laser pulses as well as
a time-independent Heisenberg interaction between each pair of electron spins.
We discuss the individual dephasing contributions stemming from the Overhauser field, 
the distribution of the electron $g$-factors and the electronic spin-spin
interaction  as well as the spectral width of the laser pulse.
This analysis reveals the counterbalancing effect of the total dephasing time when increasing the spectral laser width.
On one hand, the deviations of the electron $g$-factors increase.
On the other hand, an increasing number of coherently pumped electron spins synchronize due to the spin-spin interaction.
We find an excellent agreement between the experimental data and the dephasing time
in the simulation using an exponential distribution of Heisenberg couplings with a
mean value $\overline{J}\approx \SI{0.26}{\mu eV}$.

\end{abstract}
\maketitle

\section{Introduction}

The implementation of quantum information technologies in solid state systems was
triggered by potential advantages such as integrability, robustness and
scalability.
Focusing on semiconductors has appeared particularly appealing because of
established technology platforms to fabricate devices and the possibility to connect them to classical information processing components.
Here, carrier spins in quantum dots (QDs) have been identified to be
possible quantum bit candidates
\cite{GreilichOulton2006,Hanson2007,Greilich2007}, because the rather
efficient spin relaxation in semiconductors is mostly associated with free carrier
motion which is suppressed by the three-dimensional confinement in dot structures.
The electron-nuclear hyperfine interaction limits the electron spin coherence in
these QDs, but the decoherence due to the resulting Overhauser field
\cite{Merculov2002} can be partially reduced \cite{GreilichOulton2006,Greilich2007}
by periodic optical pulses that induce a synchronization between the electron and
the nuclear spin dynamics \cite{Urbaszek2013}. 

Manipulations on single QDs are often limited to an electrical readout
\cite{Elzerman04,Bechtold2015} of the spin-state that is perturbative.
Non-perturbative measurements by a weak probe laser pulse require a substantial
optical signal provided by an ensemble of QDs
\cite{GreilichOulton2006,Greilich2007}.
Due to the large average distance of the QDs, the dephasing measured on this
ensemble should be determined by the sum of the individual contributions.
Two-color pump-probe experiments \cite{Spatzek2011}, however, clearly revealed the
existence of an effective spin-spin coupling between the electron spins of different
QDs, that is in the order of \si{\micro eV} and of still unknown microscopic origin.

In this paper, we explore the influence of tailored pulse shapes from a single pump
laser onto an ensemble of QDs in an external magnetic field applied perpendicular to
the optical axis.
A pump pulse induces a finite spin polarization that precesses around the external
magnetic field, and the polarization amplitude decays due to various dephasing
mechanisms.
Disorder in the growth process of the self-assembled QDs generates a slight
variation in the shapes and compositions of the QDs that influences the hyperfine interactions, the
effective electron $g$-factor as well as the trion excitation energy.
While a detuned trion energy strongly suppresses the pump efficiency for a single
frequency laser, the efficiency of a time-shaped pump pulse with a finite spectral
frequency width is more complex.
Here, all QDs of the ensemble with different trion energies are subject to the same
time-shaped pump pulse such that the effect of the laser pulse varies from QD to QD
and needs to be individually taken into account. 

Surprisingly, the integral effect of all dephasing contributions yields an external
magnetic field dependent dephasing time that is almost constant as a function
of the spectral laser width.
Naively, one would expect a reduction with increasing spectral width: More and more
sub-ensembles of QDs are contributing to the total electron spin polarization which
should lead to a reduction of the dephasing time due to the increasing spread of the
electron $g$-factors. 
However, experimental data presented in this paper contradicts this naive assumption and displays a dephasing time nearly independent of the spectral laser width.

We address this question by a numerical simulation of the spin dynamics of coupled
QDs for various parameter regimes.
We restrict ourselves to the dominant interactions: the hyperfine interaction
\cite{Gaudin1976,Merculov2002,Hanson2007,Urbaszek2013}, the
electronic and nuclear Zeeman energy, and the pump laser
\cite{Barnes2011,Jaeschke2017,Kleinjohann2018}.
Since the microscopic origin of an interaction between the QDs is still unresolved,
we model the effective electronic spin-spin interaction with a simple
time-independent Heisenberg term between each pair of electron spins in the system
and leave the discussion on the effect of the nuclear-electric quadrupolar
interaction and the dipole-dipole terms to a future study. 

Establishing the existence of such a long range spin-spin interaction and
determining the required magnitude of the average value might provide a basis for a
microscopic explanation.
It has been speculated \cite{Spatzek2011} that the spin-spin interaction might be
related to an optical RKKY mechanism proposed in Ref.\ \cite{Piermarocchi2002}.
This can be probably ruled out since a time-independent interaction is required to
explain the experimental data.
It could be that the doping of the QD layers leads also to a very weak doping of the
wetting layer that connects the QDs which would generate a conventional RKKY 
interaction between the electron spins of the QDs.

In order to address larger numbers of nuclear spins and a reasonable number of
coupled QDs, we resort to a semiclassical approach (SCA) for which the simulation
time scales linearly with the number of degrees of freedom.
This SCA was derived \cite{Chen2007} from a saddle-point approximation of a
path-integral formulation for the quantum mechanical spin dynamics where the quantum
mechanical trace is replaced by the integral over the Bloch sphere of each spin.
The approach is sampled with increasing accuracy by a configuration average over the
classical spins and has been extended to periodically pulsed QDs
\cite{Jaeschke2017}.
We extend the SCA to the simulation of a coupled cluster of a finite number of QDs:
the effect of the Heisenberg coupling of all other QDs leads to an additional
time-dependent magnetic field.
Since additional fluctuations remain finite due to the distance-dependent decay of the 
spin-spin interaction, it is sufficient to investigate a finite size cluster.

Focusing on the fluctuations of the time-dependent effective magnetic field acting
on the electron spin of one QD reveals already the basic competing mechanisms that
eventually lead to an almost constant dephasing time with the laser spectral width.
If that QD is optimally pumped for a very narrow spectral width, all other QDs will
be hardly affected and just provide an additional source of dephasing for the
electron spin polarization.
With increasing spectral width more and more QDs are simultaneously pumped.
This leads to a synchronization after the pump pulse and a significant reduction of
the magnetic field fluctuations. 
Nevertheless, the Larmor precession in each QD is governed by the individual
electron $g$-factor: its distribution is a source of increasing dephasing.
While this simplified picture qualitatively explains the weak dependency of the
dephasing time on the spectral width of the laser, our analysis below provides a
detailed quantitative discussion of the competing effects.

The paper is organized as follows.
In order to set the stage for the simulations, we discuss the experimental setup and
present experimental data on the dephasing time in Sec.\ \ref{sec:experiment}.
In Sec. \ref{sec:csm} we introduce the extended Gaudin \cite{Gaudin1976} or
central spin model (CSM) and review the SCA in Sec.\ \ref{sec:SCA}. 
The quantum mechanical description of an in general detuned laser pulse onto the
electron spin is addressed in Sec.\ \ref{sec:pulse}.
The first part of Sec.\ \ref{sec:effects} is devoted to discuss each dephasing
contribution individually:
the Overhauser field contribution in Sec.\ \ref{sec:TN}, the effect of the electron
$g$-factor distribution in Sec.\ \ref{sec:Tg}, and the electronic spin-spin
interaction in Sec.\ \ref{sec:TJ}.
We provide a statistical analysis and simplified toy models to illustrate the
dependence of $T^*$ on the parameters in certain limiting cases.
Finally, the results for simulations in the presence of all interaction terms are
provided in Sec.\ \ref{sec:results} for different effective spin-spin coupling
strength.
By combining the experimental data with the theoretical results, we are able to
determine a mean spin-spin coupling strength for which the best fit between theory
and experiment is obtained.
We end the paper by a short summary in Sec.\ \ref{sec:conclusion}.

\section{Experiment}
\label{sec:experiment}
  
\begin{figure}[t]
    \includegraphics[scale=1]{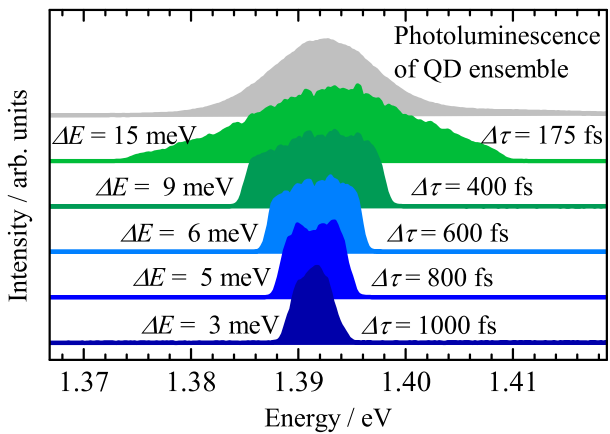}
    \caption{Laser pulse spectra with varying spectral widths $\Delta E$ and corresponding temporal lengths $\Delta \tau$. The uppermost spectrum is the photoluminescence (PL) emission of the QD ensemble.}
    \label{fig:fig1}
  \end{figure}

  The investigated sample contains 20 layers of self-assembled (In,Ga)As QDs separated by 80-nm wide GaAs barriers~\cite{GreilichOulton2006}. The dot density per layer amounts to $10^{10} \, \text{cm}^{-2}$. A Si-$\delta$-doping sheet beneath each layer provides a resident electron occupation for the dot structures. Thermal annealing of the sample shifts the band gap into the energy range both of the photon emission from Ti:Sapphire laser oscillators and significant sensitivity of Si-based avalanche photodiodes. For the studied sample, the PL emission is centered around $1.393\,$eV with a full width at half maximum (FWHM) of $13\,$meV, see the upper trace in Fig.~\ref{fig:fig1}. Previous studies indicated that at least 50\% of the dots are singly-charged while the others are neutral or doubly charged ~\cite{Spatzek2011}.

  For pump-probe Faraday- or Kerr-rotation experiments we use a Ti:Sapphire laser emitting pulses with a duration of $180\,$fs at a repetition rate of $76\,$MHz. The pulse duration corresponds to a spectral width of $15\,$meV. The central photon energy was tuned to the maximum of the PL emission. To tailor the spectral width, the pulses are diffracted at a grating. The diffracted pulses widen spatially and the beam profile shows a linear photon energy distribution. A slit is placed into the widened beam. By varying the aperture of the slit, the spectral width can be continuously reduced down to a minimum of $3\,$meV FWHM corresponding to a pulse duration of about $1\,$ps. In an alternative laser configuration, pulses with a duration of $2 \,$ps and a spectral width of $1.5\,$meV are emitted.

  The laser spectra for different spectral widths are shown in comparison to the QD emission spectrum in Fig.~\ref{fig:fig1}. Without narrowing, the pulses are broader than the emission band. With the grating-slit arrangement they can be narrowed such that only a fraction of QDs is excited. The numbers at each laser spectrum give the spectral widths and temporal durations of the pulses, respectively.
  
  The laser beam is then split into a pump and a probe beam. The circularly polarized pump pulses periodically excite a spin polarization of resident electrons along the optical axis~\cite{GreilichOulton2006}. This spin polarization subsequently precesses about an external magnetic field, applied perpendicular to the optical axis (Voigt geometry). To measure the spin polarization, the ellipticity of the initially linearly polarized probe pulses is measured after transmission through the sample. Both pump and probe beams are focused to a spot size of approximately $70 \, \mu$m on the sample, leading to the excitation and probing of $10^6$ QDs. The time evolution of the electron spin polarization is obtained by varying the time delay between the pump and probe pulse trains. To ensure a constant pump pulse area around $\Theta=\pi$, where the pump efficiency does not depend sensitively on slight variations of the pump density, the spectral pump-power density is reduced according to the spectral width such that it is about $\sim4.5\,$mW/meV. The probe beam intensity is kept always about ten times weaker than the pump.
    \begin{figure}[t]
    \includegraphics[scale=1]{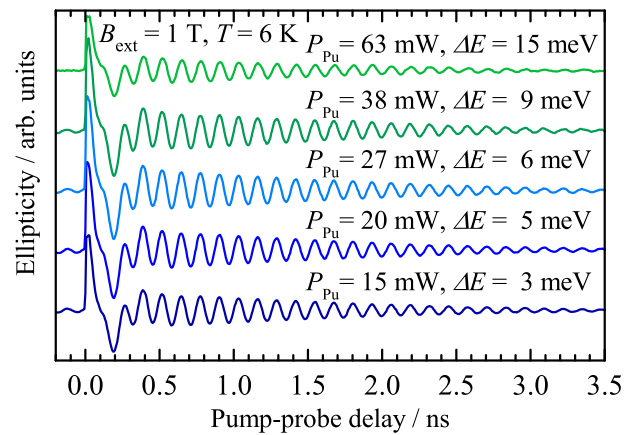}
    \caption{Time-resolved ellipticity traces showing the spin precession and dephasing of the QD ensemble at $B_\mr{ext}=1$\,T and $T=6$\,K for the different pulse widths given in Fig.~\ref{fig:fig1}.}
    \label{fig:fig2}
  \end{figure}
  \subsection{Experimental data}

  \begin{figure*}[t]
    \includegraphics[scale=1]{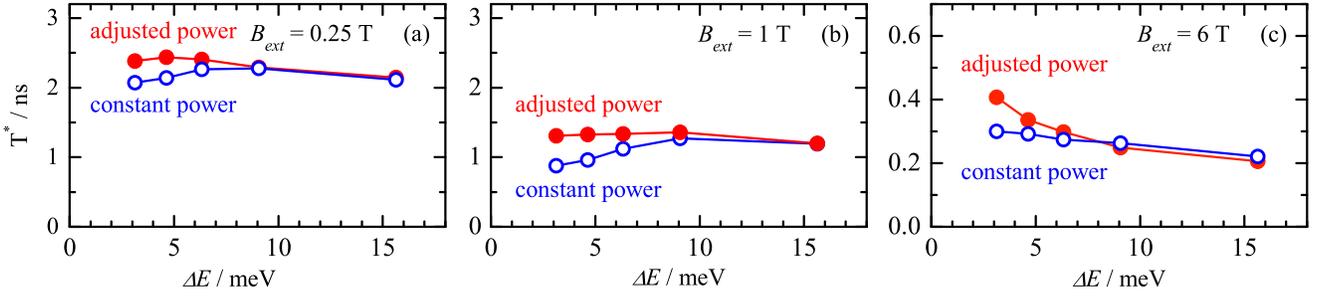}
    \caption{Measured (red, blue) signal decay times in dependence of the spectral pulse width $\Delta E$ at $B_\mr{ext}=0.25\,$T (a), $B_\mr{ext}=1\,$T (b) and $B_\mr{ext}=6\,$T (c).} \label{fig:fig4}
  \end{figure*}
  
  \begin{figure}[t]
    \includegraphics[scale=1]{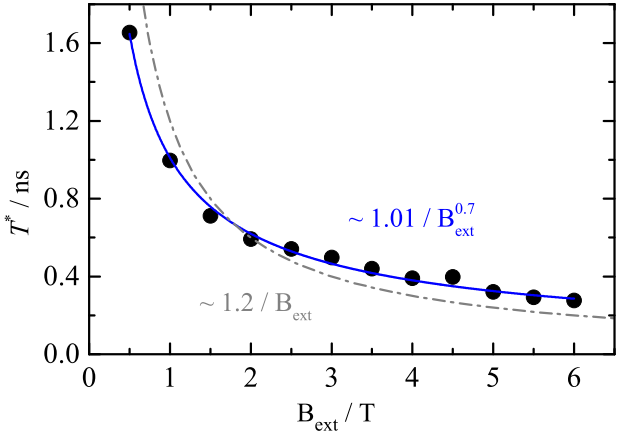}
    \caption{Signal decay time for a fixed pulse duration of $2\,$ps in dependence of the external magnetic field $B_\mr{ext}$. The expected $1/B_\mr{ext}$ dependence is depicted by the dash-dotted gray line. The blue solid line shows a fit to the data proportional to $1/B_\mr{ext}^{\alpha}$.
    } \label{fig:fig5}
  \end{figure}
  
  Figure~\ref{fig:fig2} shows pump-probe Kerr rotation traces recorded for varying spectral width of the laser pulses, as discussed above. The traces were taken at $B_\mr{ext}=1\,$T. For delays of more than $\sim 0.5\,$ns, after which the optically excited trions have decayed, the signal shows damped oscillations. The period of the oscillations is given by the average precession frequency in the excited dot ensembles corresponding to the average electron $g$-factor. Since the central photon energy is kept at the same position for the different traces, we observe the same precession frequency independent of the pulse spectral width.
  
  Also the decay which occurs on a nanosecond time scale looks very similar for the different traces. This is a very surprising finding because the underlying damping is not related to loss of the coherence of the individual spins in each QD of the ensemble. The associated coherence time has been shown to be in the microsecond range~\cite{GreilichYakovlev2006}. But it is rather an ensemble effect resulting from the inhomogeneity of the excited $g$-factor distribution. This leads to dephasing, hence the destructive interference of the contribution to the signal, arising from the distribution of corresponding precession frequencies. However, according to previous studies of the electron $g$-factor in these QDs, the width of this precession frequency distribution is expected to increase strongly with increasing spectral width of the exciting laser pulse. Consequently, the signal decay that results from this inhomogeneity should occur the faster the broader the spectral width is.
  
  In a naive approach, the characteristic time scale can be given by the $g$-factor variation $\Delta g$ in the optically excited QD ensemble 
  \begin{align}
  {T^*}^{-1} \propto \Delta \omega_{\text{L}} = \frac{\Delta
    g \mu_{\text{B}} B_\mr{ext}}{\hbar} \label{eq:Tinh}
  \end{align}
  with the average Larmor frequency $\omega_{\text{L}}$ and the spread of the Larmor frequency $\Delta \omega_{\text{L}}$ as well as with the reduced Planck constant $\hbar = 1$. The $g$-factor varies to a good approximation linearly with the optical transition energy~\cite{GreilichYakovlev2006}. The pulse spectral width $\Delta E$ varies by a factor of about 6 in our experiments so that a similar variation is expected for $\Delta g \propto \Delta E$. Accordingly the decay time is expected to decrease by a factor of 6 when changing from excitation by pulses with $1000\,$fs duration to $180\,$fs pulses. This is obviously not the case in the experimentally measured traces.
  
  We determine the signal dephasing time from a fit to the data by a Gaussian-damped cosine function:
  \begin{align}
  S_z(t)=A \cos (\omega_{\text{L}} t) \exp{\left(-\frac{t^2}{2 {T^*}^2} \right)} \, . \label{eq:TstarFit}
  \end{align}  
 The decay times are plotted against the spectral width of the laser for three different magnetic field strengths in Fig.~\ref{fig:fig4}. Panel (b) gives the data for $B_\mr{ext} = 1\,$T, where the red full dots show the results from the measurements of Fig.~\ref{fig:fig2}. As expected from the observation there, within the experimental accuracy there is no variation of the dephasing time $T^*$ with the spectral pulse width. As indicated, these studies were performed with the spectral excitation power density fixed. To test whether the observed constancy is a result of the specific excitation conditions, we have also performed measurement with the total excitation power fixed and only the pulse duration changed. These results are given by the open symbols and confirm basically the observed trend: The dependence on the spectral width is weak and even opposite to the expectations: With decreasing spectral density, the dephasing time drops to slightly below a nanosecond for $B_\mr{ext}=1\,$T.
  
  To obtain more insight into the observed dependence, we have also checked the magnetic field dependence of the dephasing time using a fixed laser spectral width of $1.5\,$meV. According to Eq.~\eqref{eq:Tinh} from above the dephasing for not too small magnetic fields is inversely proportional to the magnetic field strength, which gave a reasonable description of the trend in observed data that is also seen here from the shortening of the dephasing time with increasing $B_\mr{ext}$ in Fig.~\ref{fig:fig4}. 
  
  Fig.~\ref{fig:fig5} shows the magnetic field dependence of the signal dephasing time for a fixed pulse duration of $2\,$ps corresponding to a spectral width of about $1.5\,$meV. The dephasing time decreases from about $1.6\,$ns at $0.5\,$T to almost $0.2\,$ns at $6\,$T. A fit to the data according to a $1/B_\mr{ext}$-dependence describes this trend (the gray dash-dotted line), but has significant deviations. Much better agreement with the data is obtained, however, from a fit with variable exponent $1/B_\mr{ext}^{\alpha}$ which gives $\alpha$ = 0.7, see the solid blue line. This is another clear indication that the signal dephasing is not solely governed by the $g$-factor inhomogeneity of independent spins, but is also influenced by other factors.
  
 An obvious candidate to explain is the expected electronic spin-spin interaction in the ensemble, so far demonstrated for two distinct spin ensembles excited by different laser pulses only~\cite{Spatzek2011}. The interaction should be acting, however, between the spins in different excited spin ensembles within a spectrally broadened laser pulse as well as between the non-excited electron spins. Transferring this consideration to a situation with more than two interacting spin ensembles leads to smearing out of the modulation as the individual spin ensembles precess about each other. However, this does not represent a useful criterion for identifying the interaction between the electron spins. More conclusive could be the influence on the signal decay time, where we observed only a weak dependence on the spectral width of the laser pulses. This behavior is a result of the superposition of (i) the fluctuations of the nuclear spin bath, (ii) the inhomogeneities in the electron $g$-factor in the spin ensemble, and (iii) the electronic spin-spin interactions. In the following, we present a model which comprises and quantifies these mechanisms.

\section{Model and methods}
\label{sec:model}

In this section we present the model and the theoretical approach that is able to
describe and explain the experimental findings.
We start with the introduction of an extended central spin model \cite{Gaudin1976} that we treat with a SCA \cite{Merculov2002,Chen2007,Glazov2012,Jaeschke2017}.
We combine the spin dynamics between the laser pulses with a Lindblad approach and use a quantum mechanical description for the effect of the laser pulses.
The SCA \cite{Chen2007} overcomes the  the exponential growth of the Hilbert space since the number of degrees of freedom only grows linearly with the number of spins. The quantum mechanical trace is replaced by a configuration average. 
We use the self-averaging in case of a large number of configurations to include the ensemble averaging of different QDs.

\subsection{Extended central spin model}
\label{sec:csm}

The leading contribution to the central spin dynamics in $n$-type singly-charged semiconductor QDs is accounted for by the CSM: the central spin is coupled to a bath of nuclear spins while there is no interaction among the nuclear spins.
This assumption is justified, since the hyperfine interaction between the electron spin and the nuclear spins is several orders of magnitude  stronger than the nuclear dipole-dipole interaction or the nuclear quadrupolar interaction \cite{Dyakonov2008}.

Within the CSM, typically a single QD is described.
To model a QD ensemble, the average over a set of single QDs with different properties (e.g. $g$-factors, trion excitation energies and hyperfine coupling constants) could be performed.
But in this case the spin dynamics of the QDs would be independent of each other.
Experimentally, however, two-color pump-probe experiments \cite{Spatzek2011}
have shown that different QDs are correlated.
The coupling has  been phenomenologically identified as a time-independent long-range Heisenberg term of unknown microscopic origin.

In the present paper, we address the influence of such an interaction between the electron spins of different QDs. 
Therefore, the extended CSM comprises the contribution for each QD $i$,
\begin{align}
	H_1^{(i)} =& g_{\mr{e}}^{(i)} \mu_\mr{B} \vec{B}_\mr{ext} \vec{S}^{(i)} 
	+ \sum_{k} g_{\mr{N},k}^{(i)} \mu_\mr{N} \vec{B}_\mr{ext} \vec{I}_k^{(i)} \nonumber\\
	&+ \sum_{k} A_k^{(i)} \vec{I}_k^{(i)} \vec{S}^{(i)} \label{eq:H1},
\end{align}
including the electron and nuclear Zeeman term in the external magnetic field $\vec{B}_{ext}$
as well as the hyperfine coupling between the nuclei and the electron spin,
and the Heisenberg interaction between each pair $(i,j)$ of QDs
\begin{align}
H_2^{(i,j)} = J_{i,j} \vec{S}^{(i)} \vec{S}^{(j)} \label{eq:HI}
.
\end{align}
The total Hamiltonian is given by
\begin{align}
	H =& \sum_i H_1^{(i)} + \sum_{i,j \, (i\neq j)} H_2^{(i,j)} \label{eq:H} 
\end{align}

The indices $i,j \in \{1,...,N_\mr{QD}\}$ label the different QDs and the index $k \in \{1,...,N_\mr{N}\}$ 
denotes the different nuclear spins within a QD.
The coupling strength to the external magnetic field is given by the electron and nuclear $g$-factors $g_e^{(i)}$ and $g_{N,k}^{(i)}$ as well as the Bohr and nuclear magneton $\mu_\mr{B}$ and $\mu_\mr{N}$, respectively.
The coupling constants of the hyperfine interaction and the electronic spin-spin interaction are labeled $A_k^{(i)}$ and $J_{i,j}$.

\subsection{Semiclassical approach}
\label{sec:SCA}

To overcome the exponential growth of the Hilbert-space, we use a SCA
\cite{Jaeschke2017}
for the calculation of the time evolution after a laser pulse.
The SCA can be derived from a saddle-point approximation in the quantum mechanical path integral formulation \cite{Chen2007}. 
By introducing spin coherent states, the quantum mechanical trace is
replaced by a classical configuration average over spins on a Bloch sphere.

The decay of the trion, that is generated by the laser pulse, has to be taken into account.
The trion decay is combined with the SCA using the quantum mechanical Lindblad formalism for open quantum systems \cite{Carmichael2013,Jaeschke2017}.
This requires the trion probability $P_\mathrm{T}^{(i)}$ as an additional parameter of the system \cite{Jaeschke2017}.
Quantum mechanically, $P_\mr{T}^{(i)}$ represents the occupation number of the trion state $\ket{\mr{T}^{(i)}}=\ket{\uparrow \downarrow \Uparrow^{(i)}}$ in QD $i$, i.e. $P_\mr{T}^{(i)} = \mathrm{Tr}[\ket{\mr{T}^{(i)}}\bra{\mr{T}^{(i)}} \; \rho]$ with the density operator $\rho$.

The quantum mechanical equations of motion are replaced by their classical counter part
\begin{align}
		\frac{\mathrm{d}}{\mathrm{d}t}\vec{S}^{(i)} &= \vec{B}_\mathrm{tot}^{(i)}(t) \times \vec{S}^{(i)} +\gamma P_\mathrm{T}^{(i)} \frac{\vec{e}_z}{2} \label{eq:dS}\\
			\frac{\mathrm{d}}{\mathrm{d}t} \vec{I_k}^{(i)} &= \vec{B}_{\mathrm{tot},k}^{(i)}(t) \times \vec{I_k}^{(i)}\label{eq:dPT} \\
			\frac{\mathrm{d}}{\mathrm{d}t} P_\mathrm{T}^{(i)}  &= -\gamma P_\mathrm{T}^{(i)}(t)
			\label{eq:dI}
\end{align}
with a trion decay rate $\gamma\approx \SI{10}{ns^{-1}}$ \cite{Jaeschke2017}.

At each point in time, the equations describe the precession of each spin driven by
a time-dependent effective magnetic field that consists of the following
individual contributions:
\begin{align}
	\vec{B}_\mathrm{tot}^{(i)} &=  g_\mathrm{e}^{(i)}\, \mu_\mathrm{B}\vec{B}_\mathrm{ext} + \sum_k A_k^{(i)} \vec{I}_k^{(i)} + \sum_j J_{i,j} \vec{S}^{(j)}  \\
	\vec{B}_{\mathrm{tot},k}^{(i)} &=  g_\mathrm{N}\, \mu_\mathrm{N}\vec{B}_\mathrm{ext} + A_k^{(i)} \vec{S}^{(i)} \; .
\end{align}
For the central spin in QD $i$, this total field $\vec{B}_\mathrm{tot}^{(i)}(t)$ comprises the external magnetic field, the Overhauser field and the effective magnetic field $\vec{B}^{(i)}_J$ caused by the electron spins in the other QDs,

\begin{align}
 \vec{B}^{(i)}_J &= \sum_j J_{i,j} \vec{S}^{(j)} \; .
\end{align}
For the nuclear spin $k$ in the QD $i$, the total magnetic field 
$\vec{B}_{\mathrm{tot},k}^{(i)}$
is composed of the external magnetic field and the Knight field of the respective electron spin.

The electron spin precession is superimposed by the trion decay.
The trion probability $P_\mr{T}^{(i)}$ decays exponentially with the decay rate $\gamma$ into the central spin component along the optical axis ($z$-direction).
The only difference to the equations employed in Ref.\ \cite{Jaeschke2017} is the additional effective field $\vec{B}^{(i)}_J$ generated by the surrounding QDs.

In our numerical calculation, Eqs. \eqref{eq:dS} to \eqref{eq:dPT} are solved for $N_\mr{conf}$ 
classical configurations with different initial vectors of $\vec{S}^{(i)}$ and $\vec{I}^{(i)}_k$.
Since the configurations are independent of each other, we omit a 
potential index labeling the configurations for the sake of simplicity.
For the initial condition, we assume the thermal energy to be much higher than the energy scales of the Hamiltonian in Eq. \eqref{eq:H} but much smaller than the trion excitation energy $\epsilon_\mr{T}$.
Thus, $\vec{S}^{(i)}$ and $\vec{I}^{(i)}_k$ are randomly aligned on the surface of the Bloch sphere and the trion state is empty.

In the derivation of the SCA, the values $g_\mathrm{e}^{(i)}, A_k^{(i)}$ and $J_{i,j}$ would be the same within each classical configuration \cite{Chen2007}.
To mimic the ensemble averaging over many QDs in the sample, 
we choose $g_\mathrm{e}^{(i)}$, $A_k^{(i)}$ and $J_{i,j}$ in each configuration from the distributions $p(g_\mathrm{e}^{(i)})$, $p(A_k^{(i)})$ and $p(J_{i,j})$, respectively.

In the extended version of the CSM, the effect onto QD $i$ caused by the other QDs is given by the vector $\vec{B}^{(i)}_J$.
If the electron spins are randomly distributed and uncorrelated, $\vec{B}^{(i)}_J$ is just another Gaussian distributed
field. Instead of $N_{\rm QD}\approx10^6$, as in a real QD sample, one can generate  $\vec{B}^{(i)}_J$ with a relatively small number of QDs
and use appropriately scaled couplings $J_{i,j}$ such that the fluctuations $\langle (\vec{B}^{(i)}_J)^2 \rangle $ are identical.
The same is true for synchronized sub-ensembles of QDs: the noise can either be represented by 
all the QDs of the sub-ensemble contributing to $\vec{B}^{(i)}_J$ individually, or by choosing a 
representative spin for the sub-ensemble and include the randomization in the generation of the configurations.
Using the self-averaging, we can convince ourselves that the we can mimic the ensemble by a 
relatively small number of $N_\mr{QD}=10$ representative QDs plus configuration randomization.

\subsection{Laser pulses}
\label{sec:pulse}

Quantum mechanically, a laser pulse with $\sigma^+$-polarization creates a transition between the spin up state $\ket{\uparrow}$ and the trion state $\ket{\mathrm{T}}$ while the spin down state $\ket{\downarrow}$ is unaffected.
Here, the quantization axis matches the optical axis.
In our considerations, we neglect the trion state $\ket{\uparrow \downarrow \Downarrow}$ since we focus on $\sigma^+$-pulses.
To model the pump pulses, we start from a quantum mechanical derivation and then transform the quantum mechanical pulse action into the semiclassical picture.

In quantum mechanics, the effect of the pump laser pulse can be described by a unitary transformation.
Since the pulse is much shorter than the dynamics given by the Hamiltonian H (see Eq. \eqref{eq:H}), we omit the effects of $H$ during the laser pulse duration.
Thus, the QDs can be treated separately.
In a rotating wave approximation, the Hamiltonian of light-matter interaction is given by
\begin{align}
	H_\mathrm{L}^{(i)}(t) &= \epsilon_\mathrm{T}^{(i)} \ket{\mathrm{T}} \bra{\mathrm{T}} \\
	&+ \frac{\Omega(t)}{2} \left( \exp\left(-\mr{i} \epsilon_\mathrm{L} t\right) \ket{\mathrm{T}} \bra{\uparrow}  + \exp\left(\mr{i} \epsilon_\mathrm{L} t\right) \ket{\uparrow} \bra{\mathrm{T}} \right) \notag \; ,
\end{align}
where a classical laser field drives the dipole transition between the $\ket{\uparrow}$ and $\ket{\mathrm{T}}$ state. 
The trion excitation energy $\epsilon_\mathrm{T}^{(i)}$ is distinct for each QD and follows a distribution $p(\epsilon_\mr{T}^{(i)})$ (see Sec. \ref{sec:Tg}).
The photon energy $\epsilon_\mr{L}$ determines the fast oscillations of the electro-magnetic field.
The time-dependent Rabi frequency $\Omega(t)$ is proportional to the envelope function of the laser pulse.
By a unitary transformation into the rotating frame of the laser field
\begin{align}
	U_\mathrm{L} = \exp(-\mr{i} \epsilon_\mathrm{L} t) \ket{\mathrm{T}} \bra{\mathrm{T}} \; ,
\end{align}
we obtain the Hamiltonian
\begin{align}
	\tilde{H}_L^{(i)}(t) &= U_\mathrm{L}^\dagger \; \left(H_\mathrm{L}^{(i)}(t) - \epsilon_\mathrm{L}\ket{\mr{T}} \bra{\mr{T}}\right)\; U_\mathrm{L} \\
	&= (\epsilon_\mathrm{T}^{(i)} - \epsilon_\mathrm{L}) \ket{\mathrm{T}} \bra{\mathrm{T}} + \frac{\Omega(t)}{2} \left( \ket{\mathrm{T}} \bra{\uparrow} + \ket{\uparrow} \bra{\mathrm{T}} \right) \; ,
\end{align}
where the fast oscillations with frequency $\epsilon_\mathrm{L}$ vanish.
In this case, the unitary operator $U_P$ can be numerically calculated using a proper discretization
\begin{align}
	U_P^{(i)} &\approx U_\mathrm{L} \left( \prod_l \exp\left(-\mr{i}\, \tilde{H}^{(i)}_\mathrm{L}(t_l) \,\Delta t\right) \right) U_\mathrm{L}^\dagger \; 
\end{align}
which becomes exact in the limit $\Delta t\to 0$ and a finite pulse duration $T_p=N\Delta t =\mr{const}$ .

Since $\ket{\downarrow}$ is an eigenstate of $H_\mathrm{L}(t)$ and consequently is not affected by the pump pulse, the matrix representation of the pulse operator has the form
\begin{align}
U_\mathrm{P}^{(i)} = \begin{pmatrix}
a^{(i)} & 0 & b^{(i)}\\
0 & 1 & 0\\
c^{(i)} & 0 & d^{(i)}\\
\end{pmatrix} \label{eq:UP}
\end{align}
in the basis $\ket{\uparrow}$, $\ket{\downarrow}$ and $\ket{T}$. 
Here, $a^{(i)}, b^{(i)}, c^{(i)}$ and $d^{(i)}$ are complex numbers, that are constrained by the fact that $U_\mathrm{P}^{(i)}$ has to be unitary.
The exact value of these four parameter depends on the detuning ${\delta^{(i)} = \epsilon_\mr{L}-\epsilon_\mr{T}^{(i)}}$ and the shape $\Omega(t)$ of the laser pulse. 
The detuning $\delta^{(i)}$ between the photon energy and trion excitation energy is different for each QD.

As a next step, the quantum mechanical pulse operator $U_\mathrm{P}^{(i)}$ has to be translated into the semiclassical picture.
By using the correspondence principle, we obtain the relation
\begin{align}
O_\mathrm{ap}^{(i)} = \mathrm{Tr}\left[U_p^{(i)} \rho_\mathrm{bp}^{(i)} U_p^{(i) \dagger} \hat{O}^{(i)} \right] \; , \label{eq:Oap}
\end{align}
where $O_\mathrm{ap}^{(i)}$ is the semiclassical variable 
($S_x^{(i)}, S_y^{(i)}, S_z^{(i)}$ or $P_\mathrm{T}^{(i)}$) after the pulse and 
$\hat{O}^{(i)}$ is the corresponding quantum mechanical operator \cite{Jaeschke2017} .
Assuming that the trion population has completely decayed before the pulse, the density matrix $\rho_\mathrm{bp}^{(i)}$ is given by
\begin{align}
\rho_\mathrm{bp}^{(i)} = \begin{pmatrix}
\frac{1}{2}+S_{z,\mathrm{bp}}^{(i)} & S_{x,\mathrm{bp}}^{(i)} + \mr{i} S_{y,\mathrm{bp}}^{(i)} & 0\\
S_{x,\mathrm{bp}}^{(i)} - \mr{i} S_{y,\mathrm{bp}}^{(i)} & \frac{1}{2}-S_{z,\mathrm{bp}}^{(i)} & 0\\
0 & 0 & 0\\
\end{pmatrix} \;. \label{eq:rhobp}
\end{align}
Combining Eqs. \eqref{eq:UP} to \eqref{eq:rhobp}, the effect of a pulse with arbitrary shape in the semiclassical picture results in
\begin{align}
S_{x,\mathrm{ap}}^{(i)} &= |a^{(i)}| \left(S_{x,\mathrm{bp}}^{(i)} \cos( \varphi^{(i)} ) + S_{y,\mathrm{bp}}^{(i)} \sin( \varphi^{(i)} ) \right) \label{eq:pulse_Sx}\\
S_{y,\mathrm{ap}}^{(i)} &= |a^{(i)}| \left(-S_{x,\mathrm{bp}}^{(i)} \sin( \varphi^{(i)} ) + S_{y,\mathrm{bp}} \cos( \varphi^{(i)} ) \right)\label{eq:pulse_Sy}\\
S_{z,\mathrm{ap}}^{(i)} &= -\frac{1}{4} \left( 1-|a^{(i)}|^2 \right) + \frac{S_{z,\mathrm{bp}}^{(i)}}{2} \left( 1+|a^{(i)}|^2  \right)\label{eq:pulse_Sz}\\
P_\mathrm{T,ap}^{(i)} &= \left(1 - |a^{(i)}|^2 \right) \left(\frac{1}{2} + S_{z,\mathrm{bp}}^{(i)}\right) \; .\label{eq:pulse_PT}
\end{align}

Note that the effect of the pulse can be parametrized by a single complex number ${a^{(i)} = |a^{(i)}| \exp(\mr{i}\varphi^{(i)})}$.
The parameters $b^{(i)}$ and $d^{(i)}$ do not enter 
since no trion is present before the pulse.
Furthermore, we used the fact, that $U_\mathrm{P}^{(i)}$ is unitary, to eliminate the parameter $c^{(i)}$.
The effect of the pulse, given by Eqs. \eqref{eq:pulse_Sx} to \eqref{eq:pulse_PT}, and the role of the parameter $a^{(i)}$ can be interpreted in a simple manner.
$(1-|a^{(i)}|^2)$ is the probability to excite an electron in spin up state to the trion state and is a measure for the pulse efficiency. 
The geometrical phase $\varphi^{(i)}$ rotates the spin around the optical axis and can be used e.g. for spin echo experiments \cite{Economou2006,Economou2007,GreilichEconomou2009}
or optical spin tomography \cite{Varwig2014}.
Eqs. \eqref{eq:pulse_Sx} to \eqref{eq:pulse_PT} describe the general behavior of a short laser pulse with arbitrary detuning $\delta^{(i)}$ and pulse area $\Theta = \int \Omega(t) \mr{d}t$.
In the case of resonant pulses ($\delta^{(i)}=0$), the rotation around the optical axis vanishes ($\varphi=0$) and the pulse equations from Ref. \cite{Glazov2012} result.
For resonant $\pi$-pulses ($\delta^{(i)}=0$, $\Theta=\pi$), the parameter $a^{(i)}$ yields zero and the pulse equations match those in Ref. \cite{Jaeschke2017}.
In the present work, we employ Gaussian pulses
\begin{align}
\Omega(t) = \frac{\Theta}{\sqrt{2 \pi \sigma_t^2}} \exp\left(-\frac{t^2}{2 \sigma_t^2}\right) 
\end{align}
with $\Theta=\pi$ unless stated otherwise.
The temporal pulse width $\sigma_t$ is the inverse of the spectral laser width $\sigma_E$.
The spectral FWHM is given by
\begin{align}
\Delta E = 2 \sqrt{2 \ln 2} \; \sigma_E = \frac{2 \sqrt{2 \ln 2}}{\sigma_t} \; .
\end{align}

In addition to the pump pulses as discussed above, we have to consider the probe pulse.
In the experiments, the $S_z^{(i)}$ component of the spin polarization is measured by means of the Faraday ellipticity.
The Faraday ellipticity is proportional to the trion excitation probability of the probe pulse.
In our simulation, we reproduce this behavior by weighting the $S_z^{(i)}$ component of each QD with the trion excitation probability $(1-|a^{(i)}|^2)$.
Hence, QDs with a trion excitation energy close to the probe laser energy contribute more efficiently to the signal than detuned QDs \cite{Yugova2009, Glazov2010}.
To match the experiment in Sec. \ref{sec:experiment}, we assume  that the probe laser has always the same spectral width as the pump laser.

\section{Contributions to the electron spin dephasing}
\label{sec:effects}

The dephasing of electron spins in singly-charged semiconductor QDs originates from three major effects, the nuclear spin fluctuations, the electron $g$-factor dispersion and the electronic spin-spin interaction.
These contributions exhibit a different dependency on the external magnetic field and the spectral width of the laser pulse.
We discuss the dephasing behavior for each effect separately.
Afterwards, we combine all effects and compare the results to the experimental measurements in Sec.\ \ref{sec:experiment}.
We restrict ourselves to a single laser pulse and the succeeding time evolution.
As last aspect, we analyze the effects that arise from periodically pumping the system.

\subsection{Dephasing due to the nuclear spin fluctuations}
\label{sec:TN}

\begin{figure}[t]
\begin{center}
\includegraphics{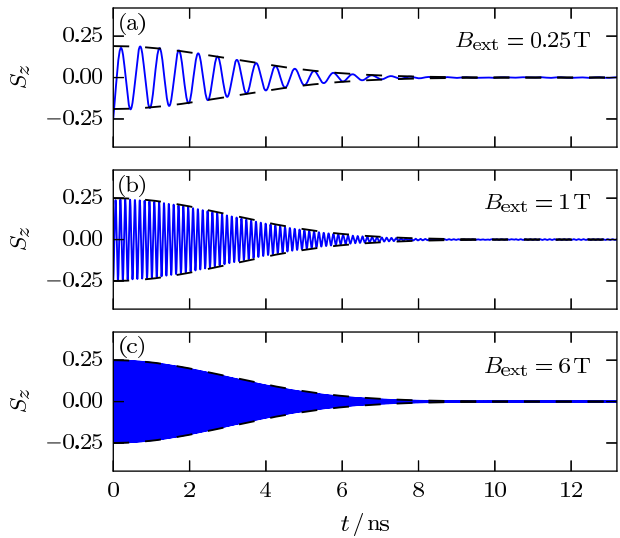}
\caption{Dephasing due to the nuclear spin fluctuations. We set the parameters to $T_\mr{N}^*=3 \, \si{ns}$, $g^{(i)} = 0.555$ and $J_{i,j}=0$. The $S_z$ component of the electron spin as function of the time $t$ after a resonant $\pi$-pulse is depicted as a blue curve for three different magnetic fields in (a), (b) and (c).
Differences between homogeneous and exponentially distributed coupling constants can not be resolved on this time scale. The dashed lines mark the analytical envelope functions according to Eq. \eqref{eq:gauss}. }
\label{fig:TN}
\end{center}
\end{figure}

At first, we focus on the influence of the nuclear spin fluctuations on the dephasing time of the electron spin.
By the hyperfine interaction, each electron spin is coupled to its own nuclear spin bath.
Thus, the electron spin $\vec{S}^{(i)}$ is affected by the Overhauser-field
\begin{align}
\vec{B}_N^{(i)} = \sum_k A_k^{(i)} \vec{I}_k^{(i)} \; ,
\end{align}
and the nuclear spins in turn are subject to the Knight field
\begin{align}
\vec{B}_k^{(i)} =  A_k^{(i)} \vec{S}^{(i)} \;  .
\end{align} 

For $N_\mr{N}$ unpolarized nuclear spins, the Knight field is by a factor $\sqrt{N_\mr{N}}$ smaller than the Overhauser field.
In real QDs with $N_\mr{N}=10^5$ nuclear spins, the Knight field is therefore several orders of magnitude smaller.
Hence, the Overhauser field can be considered as frozen for short time scales \cite{Merculov2002}.
For an unpolarized frozen Overhauser field and a strong external magnetic field ($B_\mathrm{ext}\gg B_\mathrm{N}$), it was shown \cite{Merculov2002} that the Larmor oscillation of the electron spin polarization $S_z$ dephases with a Gaussian envelope function
\begin{align}
	S_\mathrm{env}(t) = \pm S_0 \exp\left( - \frac{t^2}{2 T_\mathrm{N}^{*2}} \right) \label{eq:gauss} \; .
\end{align}
The nuclear dephasing time $T_\mathrm{N}^*$ is determined by the fluctuation of the Overhauser field
\begin{align}
	T_\mathrm{N}^{*} = \sqrt{\frac{3}{\sum_k A_k^{2} 
	\langle I_k^{2} \rangle}} \; 
\end{align}
where $\langle I_k^{2} \rangle$ is the averaged square of the spin length. 

According to the central limit theorem, for a large number of nuclei, the Overhauser field $\vec{B}_\mr{N}$ is Gaussian distributed independent of the distribution $p(A_k^{(i)})$.
The generic effect of the nuclear spin bath on the electron spin is encoded in the nuclear dephasing time $T_\mr{N}^*$ that determines the width of the Overhauser field distribution.
We use exponentially distributed hyperfine coupling constants
\begin{align}
	p(A_k^{(i)}) = \frac{1}{\overline{A}}\exp \left( - \frac{A_k^{(i)}}{\overline{A}}\right)
\end{align} 
with the mean value $\overline{A}$.
Since the details of the distribution $p(A_k^{(i)})$ do not display on short time scales as considered in the present calculations, we compare the results to homogeneous coupling constants $A_k^{(i)}=A$ (Box model \cite{Khaetskii2003}).
Note that for pulse sequences of several thousands of pulses, the details of the distribution $p(A_k^{(i)})$ may have an influence \cite{Jaeschke2017,Hackmann2014}.
In Fig.\ \ref{fig:TN}, the dephasing of the electron spin component $S_z$
due to the nuclear spin fluctuations is depicted
for three different external magnetic fields.
Here, we fix the electron $g$-factor for all QDs (and all configurations) to the same value $0.555$ based on experimental measurements \cite{GreilichYakovlev2006} and set the electronic spin-spin interaction to zero ($J_{i,j} =0$).
We consider QDs with $N_\mr{N}=100$ nuclei of length $I=3/2$.
We restrict ourselves to one sort of nuclei with an effective nuclear $g$-factor such that $g_\mathrm{N}\mu_\mathrm{N}$ is $\sfrac{1}{800}$ smaller than the electronic value $g_\mathrm{e}\mu_\mathrm{B}$
\cite{Beugeling2016,Beugeling2017,Jaeschke2017}.
The hyperfine coupling constants $A_k^{(i)}$ are normalized to the nuclear dephasing time $T_\mr{N}^*=\SI{3}{ns}$ according to experimental data \cite{GreilichYakovlev2006,Greilich2007}.
In Fig.\ \ref{fig:TN}, differences between homogeneous and exponentially distributed hyperfine coupling constants can not be resolved.
If not stated otherwise, exponentially distributed hyperfine coupling constants are used in the following.
Note that we assume all QDs to have the same nuclear dephasing time $T_N^*$.
Hence, $T_\mr{N}^*$ is independent on both, the external magnetic field and the spectral width of the laser pulse.
In real QD samples, the nuclear dephasing time strongly depends on the growth parameters and can vary an order of magnitude between different samples. 
For the sample in Sec. \ref{sec:experiment}, the Overhauser field fluctuations are roughly \SI{7.5}{mT}  \cite{Schwan2011}.

\subsection{Dephasing due to the electron $g$-factor dispersion}
\label{sec:Tg}

As a second contribution to the electron spin dephasing, we investigate the effect of the electron $g$-factor dispersion.
Due to the growth process of self-assembled QDs, inhomogeneities in the ensemble occur.
Especially the electron $g$-factor $g_\mr{e}^{(i)}$ and the trion excitation energy $\epsilon_\mr{T}^{(i)}$ vary within the ensemble of QDs.
The different electron $g$-factors lead to different Larmor frequencies at a fixed external magnetic field and consequently to a dephasing of the electron spin polarization.
First, we present the correlated distributions of $g_\mr{e}^{(i)}$ and $\epsilon_\mr{T}^{(i)}$ that we employ in our calculations and analyze the magnetic field dependency of the dephasing time.
Then, we investigate the non-linear relation between the laser band width and the dephasing time, which arises due to the correlations between the electron $g$-factor and the trion excitation energy.

\subsubsection{Distribution of the electron $g$-factors}

\begin{figure}[t]
\begin{center}
\includegraphics[]{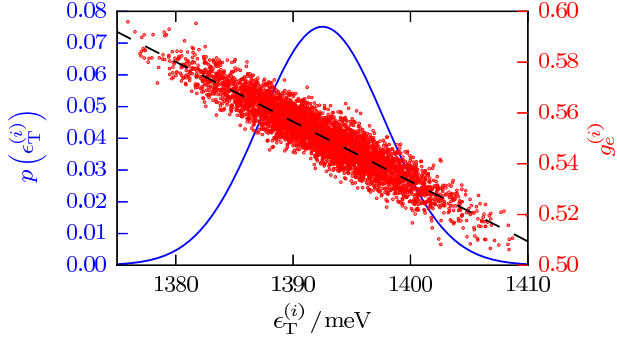}
\caption{
Distributions of the trion excitation energy $\epsilon_\mr{T}^{(i)}$ and the electron $g$-factor $g_\mr{e}^{(i)}$.
The trion excitation energy is Gaussian distributed $\epsilon_\mr{T}^{(i)} \sim \mathcal{N}(\epsilon_{\mr{T},0},\,\sigma_{\epsilon_\mr{T}})$ (solid blue line) based on the experimental photoluminescence spectrum (cf. Fig. \ref{fig:fig1}).
The linear relation between the average $g$-factor $\overline{g}^{(i)}_\mathrm{e}$ and the trion excitation energy $\epsilon_\mr{T}^{(i)}$ as stated in Eq. \eqref{eq:averageg} is shown as a black dashed line.
The red dots depict 5000 pairs $(\epsilon_\mr{T}^{(i)}, g_\mr{e}^{(i)})$ with $g_\mr{e}^{(i)} \sim \mathcal{N}(\overline{g}^{(i)}_\mathrm{e},\sigma_{g,0})$ generated randomly.}
\label{fig:gFeT}
\end{center}
\end{figure}

From the photoluminescence spectra of the self-assembled QD ensemble (cf. Fig.\ \ref{fig:fig1}), we extract a Gaussian distribution of the trion excitation energy $\epsilon_\mr{T}$
\begin{align}
p(\epsilon_\mr{T}^{(i)}) = \frac{1}{\sqrt{2 \sigma_{\epsilon_\mr{T}}^2}}
 \exp\left( -\frac{
(\epsilon_\mr{T}^{(i)} -\epsilon_{\mr{T},0})^2}{2 \sigma_{\epsilon_\mr{T}}^2}\right)
\end{align} 
with a mean value $\epsilon_{\mr{T},0}=\SI{1392.5}{meV}$ and a standard deviation $\sigma_{\epsilon_\mr{T}}=\SI{5.3}{meV}$.
Due to the Roth-Lax-Zwerdling relation, the trion excitation energy of the QD is 
linearly related to the mean electron $g$-factor \cite{Schwan2011,Yugova2009}
\begin{align}
\overline{g}(\epsilon_\mr{T}^{(i)}) = m \cdot \epsilon_\mr{T}^{(i)} + b \; .
\label{eq:averageg}
\end{align}
The parameters $m=\SI{-2.35}{eV^{-1}}$ and $b=3.83$ can be extracted from spectral dependent measurements of the electron $g$-factor \cite{GreilichYakovlev2009}.
The electron $g$-factor $g_e^{(i)}$ is assumed to be Gaussian distributed

\begin{align}
p(g_e^{(i)}) = \frac{1}{\sqrt{2 \sigma_{g,0}^2}} 
\exp\left( -\frac{(g_e^{(i)} -\overline{g}(\epsilon_{\mr{T}}^{(i)}))^2}{2 \sigma_{g,0}^2}\right)
\end{align} 
 with the mean value $\overline{g}(\epsilon_{\mr{T}}^{(i)})$ depending on the trion energy and a standard deviation $\sigma_{g,0} = 0.005$.
In Fig.\ \ref{fig:gFeT}, the distributions of $\epsilon_\mr{T}^{(i)}$ and $g_e^{(i)}$, that are employed in our calculations, are visualized. 
In numerical calculations, up to $N_\mr{conf}\cdot N_{QD} = 10^6$ pairs ($\epsilon_\mr{T}^{(i)}$, $g_\mr{e}^{(i)}$) are generated from these Gaussians.

\begin{figure}[t]
\begin{center}
\includegraphics[]{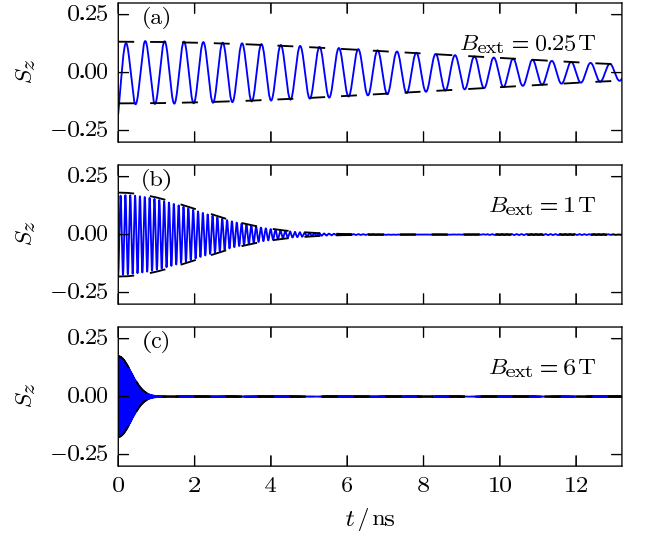}
\caption{Dephasing due to the electron $g$-factor dispersion. 
The spin dynamic with the parameters $A_k^{(i)}=0$, $J_{i,j}=0$, $\Delta E =  \SI{5}{meV}$ is depicted for three different external magnetic fields in (a), (b) and (c). The dashed lines mark the Gaussian envelope functions with standard deviation $T_g^*$ according to Eq. \ref{eq:Tg}.}
\label{fig:TG}
\end{center}
\end{figure}

To extract the parameter $\sigma_{g,0}$, we assumed that at high magnetic fields the electron Zeeman effect dominates the other contributions to the dephasing, i.e.\ the nuclear spin fluctuations or the electronic spin-spin interaction.
In this way, the deviation $\sigma_{g,0}$ is obtained from the 
measurement in Fig.\ \ref{fig:fig4} at the largest magnetic field ($B_\mathrm{ext}=\SI{6}{T}$).
In Sec. \ref{sec:results}, we 
elaborate on this point in more detail.

\begin{figure}
\begin{center}
\includegraphics[]{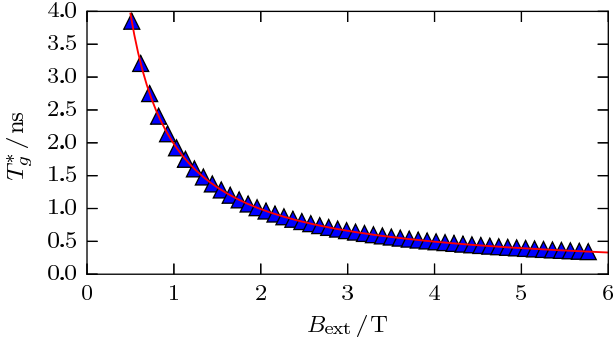}
\caption{Dependency of the electron $g$-factor induced dephasing time $T_g^*$ on the external magnetic field $B_\mathrm{ext}$. The data points (blue triangles) are extracted from numerical calculations with the parameters $A_k^{(i)}=0$, $J_{i,j}=0$, $\Delta E = \SI{5}{meV}$. The red line depicts the dependency $\propto B_\mr{ext}^{-1}$ according to Eq. \ref{eq:Tg}.}
\label{fig:Bdep}
\end{center}
\end{figure}

In Fig.\ \ref{fig:TG}, the dephasing of the electron spin polarization due to the electron $g$-factor dispersion is depicted.
The hyperfine interaction and the electronic spin-spin interaction are switched off ($A_k^{(i)}=J_{i,j} = 0$).
The dephasing is slower for weak external magnetic fields and faster for strong external magnetic fields.
The dephasing time $T_g^*$ is determined by the standard deviation $\sigma_\omega$ of the electron Larmor frequencies
\begin{align}
T_g^* = \frac{1}{\sigma_\omega} = \frac{1}{\sigma_g \mu_\mathrm{B} B_\mathrm{ext}} \; , \label{eq:Tg}
\end{align}
where $\sigma_g$ denotes the standard deviation of the electron $g$-factors.
Eq.\ \eqref{eq:Tg} is derived from the fact that the Fourier transformed of a Gaussian function is a Gaussian function with the inverse standard deviation.
The dependency $T_g^* \propto B_\mr{ext}^{-1}$ is demonstrated in Fig. \ref{fig:Bdep}, where the numerical model is compared to the analytical expression in Eq. \eqref{eq:Tg}.
The dephasing times are extracted from the numerical simulations by fitting a Gaussian envelope function (see Eq.\ \eqref{eq:TstarFit}) to the calculated time evolution.

\subsubsection{Contributions from the spectral width $\Delta E$ of the laser}

As a next step, we analyze the dependency of the dephasing time on the spectral width $\Delta E$ of the laser pulse.
Due to the correlation between the trion excitation energy and the electron $g$-factor, the deviations $\sigma_g$ grows for pulses with larger spectral width $\Delta E$.
Thus, an increase of $\Delta E$ will
lead to a faster dephasing time.

\begin{figure}
\begin{center}
\includegraphics[]{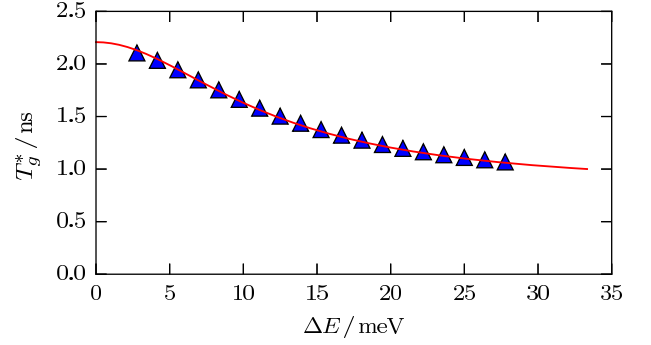}
\caption{Dependency of the electron $g$-factor induced dephasing time $T_g^*$ on the laser band width $\Delta E$. The data points (blue triangles) are extracted from numerical calculations with the parameters $A_k^{(i)}=0$, $J_{i,j}=0$ and $B_\mr{ext} = \SI{1}{T}$. The analytical solution from Eq. \eqref{eq:Tgfull} is depicted by a red line.}
\label{fig:Sigdep}
\end{center}
\end{figure}

In Fig.\ \ref{fig:Sigdep}, the dependency of the dephasing time $T_g^*$ on the spectral laser width $\Delta E$ is depicted.
For smaller spectral widths, $T_g^*$ is limited to an upper bound because of the intrinsic $g$-factor deviation $\sigma_{g,0}$.
For larger spectral widths, $T_g^*$ is restricted to a lower bound since for $\Delta E \gg \sigma_{\epsilon_\mr{T}}$ all QDs are excited by the pulse and a further expansion of $\Delta E$ has no additional effect.

In between, the value of the $g$-factor deviation $\sigma_g$ and thus the dephasing time $T_g^*$ depends on the laser band width $\Delta E = 2 \sqrt{2 \ln 2} \; \sigma_E$, the slope $m$, the deviation $\sigma_{\epsilon_{\mr{T}}}$ of the excitation energies in the ensemble and the intrinsic $g$-factor width $\sigma_{g,0}$.
The exact dependency of the dephasing time $T_g^*$ on the above stated parameters will be deduced in the following.

The $g$-factor spread $\sigma_g$
\begin{align}
\sigma_g^2 = \sigma_{g,0}^2 + \sigma_{g \Delta}^2
\end{align}
is composed of the intrinsic $g$-factor width $\sigma_{g,0}$ and a contribution $\sigma_{g \Delta} = |m|\sigma_{\epsilon_{\mr{T}\Delta}}$ that originates from the fact that the signal comprises contributions from QDs with different trion excitation energies $\epsilon_\mr{T}^{(i)}$.

To calculate the total standard deviation $\sigma_{\epsilon_{\mr{T}\Delta}}$, we introduce the weight $w(\epsilon_\mr{T}^{(i)})$ that indicates in which proportions the signal comprises contributions from the different trion excitation energies.
The weight
\begin{align}
	w(\epsilon_\mr{T}^{(i)}) &\propto p(\epsilon_\mr{T}^{(i)}) \cdot \tilde{\Omega}^2(\epsilon_\mr{T}^{(i)}) \cdot \tilde{\Omega}^2(\epsilon_\mr{T}^{(i)})
	\label{eq:weightep}
\end{align}
consists of the probability $p(\epsilon_\mr{T}^{(i)})$ to draw a QD with excitation energy $\epsilon_\mr{T}^{(i)}$ within the ensemble, the strength of the pump excitation that is proportional to $\tilde{\Omega}^2(\epsilon_\mr{T}^{(i)})$ and the probability to be measured by the probe pulse that is also proportional to $\tilde{\Omega}^2(\epsilon_\mr{T}^{(i)})$.
Here, $\tilde{\Omega}(\epsilon_\mr{T}^{(i)})$ is the Fourier transformed of the laser envelope function $\Omega(t)$.
Eq. \eqref{eq:weightep} yields
\begin{align}
w(\epsilon_\mr{T}^{(i)}) &\propto  \exp\left( -\frac{( \epsilon_\mr{T}^{(i)} - \epsilon_{\mr{T},0})^2}{2 \sigma_{\epsilon_\mr{T}}^2} \right) \cdot \exp\left( -\frac{( \epsilon_\mr{T}^{(i)} -\epsilon_{\mr{T},0})^2}{2 \sigma_E^2} \right)^4 \notag \\
	&= \exp\left( -\frac{( \epsilon_\mr{T}^{(i)} - \epsilon_{\mr{T},0} )^2}{2 \sigma_{\epsilon_{\mr{T}\Delta}}^2} \right)
\end{align}
with the total deviation
\begin{align}
	\sigma_{\epsilon_{\mr{T}\Delta}}^{-2} = \sigma_{\epsilon_\mr{T}}^{-2} + \left( \frac{\sigma_E}{2}  \right)^{-2}
\end{align}
All combined, we obtain the relation
\begin{align}
	T_g^* = \frac{1}{\mu_B B \sqrt{\sigma_{g,0}^2 + m^2 (\sigma_{\epsilon_\mr{T}}^{-2} + (\frac{\sigma_E}{2})^{-2})^{-1}}} \label{eq:Tgfull}
\end{align}
which is added as a solid (red) line to Fig.\ \ref{fig:Sigdep}.
 This relation excellently agrees with $T_g^*$ extracted from the fit to the numerical simulations.
Note that without the intrinsic $g$-factor width $\sigma_{g,0}$, the dephasing time $T_g^*$ would diverge for spectrally narrow pulses ($\sigma_E\rightarrow0$).

The two dephasing mechanisms discussed so far, i.e. the nuclear fluctuations and the electron $g$-factor spread, are independent of one another.
As a result, the nuclear dephasing time $T_\mr{N}^*$ and the $g$-factor dephasing time $T_g^*$ can be combined to a total dephasing time $T_\mr{tot}^*$ via
\begin{align}
	\left(T_\mr{tot}^* \right) ^{-2} = \left(T_\mr{N}^*\right)^{-2} + \left(T_g^*\right)^{-2} \; .
\end{align}
Before we consider the combination of dephasing mechanisms in more detail, we focus on the influence of the electronic spin-spin interaction in the next section.

\subsection{Dephasing due to the electronic spin-spin interaction}
\label{sec:TJ}

As a third contribution to the electron spin dephasing, we investigate the effect of the electronic spin-spin interaction.
An optically aligned electron spin interacts with the electron spins  in the other QDs due to a spin-spin interaction of unknown microscopic origin \cite{Spatzek2011}. 
It turns out that the experimental findings are compatible with a time-independent long-range Heisenberg term between each pair of QDs as introduced in Eq.\ \eqref{eq:HI}.
The effect of the spin-spin interaction onto the dephasing time depends strongly 
on whether the electron spins are optically aligned  or unpolarized.
To distinguish both cases, we investigate two limits of (i) spectrally narrow laser pules, where most of the QDs in the ensemble are unpolarized and (ii) spectrally broad laser pulses, where most of the QDs in the ensemble are 
affected by the optical excitation.

\subsubsection{Spectrally narrow laser pulse}

\begin{figure}[t]
\begin{center}
\includegraphics[]{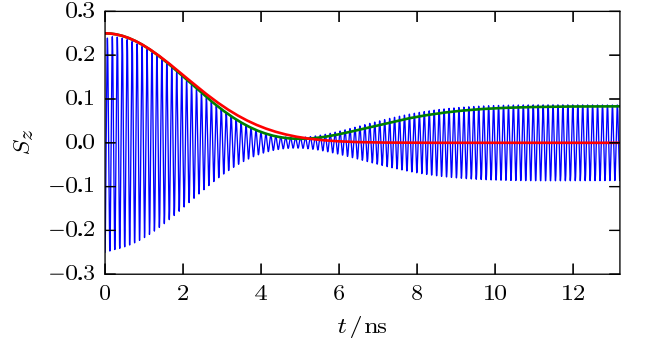}
\caption{
Dephasing due to the electronic spin-spin interaction. The blue line depicts the calculated spin dynamics with the parameters $A_k^{(i)}=0$, $g=0.555$, $\overline{J}=\SI{0.4}{ns^{-1}}$ and $B_\mathrm{ext} = 1 \, \mathrm{T}$.
The envelope function (green line) is given by Eq. \eqref{eq:mercB0}.
The initial dephasing after the pulse can be approximated by a Gaussian envelope function (red line) with standard deviation $T_J^*$.
}
\label{fig:TJ}
\end{center}
\end{figure}

In case of a spectrally narrow laser pulse, 
only a few electron spins are polarized by the laser pulse while most spins remain randomly oriented.
In Fig.\ \ref{fig:TJ}, the dynamics of interacting QDs is depicted where nuclear spin fluctuations and the electron $g$-factor dispersion are switched off. 
One of the $N_\mr{QD}=10$ QDs is pulsed with a resonant $\pi$-pulse ($a^{(1)}=0$) while the other QDs are unaffected by the laser pulse ($a^{(i)}=1$, $i=2,...,10$).
Since the dependency of the coupling strength $J_{i,j}$ on the distance $r_{i,j}$ between 
the QDs is unknown, we assume the coupling constants to be exponentially distributed
\begin{align}
	p(J_{i,j}) = \frac{1}{\overline{J}} \exp\left( - \frac{J_{i,j}}{\overline{J}} \right)
\end{align}
with a mean value $\overline{J}$.

In analogy to the Overhauser field, each QD experiences a random magnetic field $\vec{B}^{(i)}_J = \sum_j J_{i,j} \vec{S}^{(j)}$ of the other QDs.
But in contrast to the Overhauser field, $\vec{B}^{(i)}_J$ precesses in the external magnetic field with nearly the same precession frequency as the electron spin $\vec{S}^{(i)}$.
Assuming that all electrons have the same $g$-factor, in the rotating frame of the external magnetic field, neither $\vec{S}^{(i)}$ nor $\vec{B}^{(i)}_J$ precess in the external magnetic field.
Consequently, the solution in the rotating frame is given by the analytical solution in Ref. \cite{Merculov2002} at zero magnetic field
\begin{align}
S_z(t) =& \frac{S_0}{3} \Biggl( 1 + 2 \left[1 - 2 \,\frac{t^2}{2 \,T_\mr{J}^{* 2}}\right] \cdot \exp\left[ - \frac{t^2}{2 \,T_\mr{J}^{* 2}} \right] \Biggr) \label{eq:mercB0}
\end{align}
with a dephasing time
\begin{align}
T_\mr{J}^* = \sqrt{\frac{6}{\sum_j J_{i,j}^2 S_j^2}} \; .
\end{align}
Here $j$ is the index of the unpolarized QDs.
In the static frame, Eq.\ \eqref{eq:mercB0} corresponds to the envelope function of the electron spin precession.
The envelope decreases to nearly zero and rises afterwards to a third of the initial spin polarization.

The first decrease of the envelope function is described by a Gaussian envelope function with standard deviation $T_J^*$.
Consequently, we denote $T_J^*$ as the dephasing time due to the electronic spin-spin interaction for the limit of narrow pump pulses.
Note the factor $\sqrt{2}$ between $T_\mr{N}^*$ and $T_J^*$ that is caused by the fact, that $\vec{B}^{(i)}_J$ precesses with nearly the same frequency as the electron spin $\vec{S}^{(i)}$ while the Overhauser field is approximately frozen.

The total dephasing time, that has contributions from the nuclear spin fluctuations, the electron $g$-factor dispersion and the fluctuations of unpolarized electron spins, can approximately be summarized to 
\begin{align}
(T^*_\mr{tot})^{-2} \approx (T_N^*)^{-2}+ (T_g^*)^{-2} +(T_J^*)^{-2} \; .
\end{align}
This relation is valid only if the QDs are independent of each other, i.e. if only a few QDs are polarized or if the interaction between the QDs is weak.

\subsubsection{Spectrally broad laser pulse}

In the second case, most of the QDs are polarized and the electronic spin-spin interaction has a different effect.
Directly after the pulse, the effective magnetic field $\vec{B}_J^{(i)}$ is aligned along the optical axis and has no effect.
As soon as the electron spins acquire a relative phase shift due to other effects, 
$\vec{B}_J^{(i)}$ and $\vec{S}^{(i)}$ are not parallel any more.
If $\vec{S}^{(i)}$ precesses slower than $\vec{B}_J^{(i)}$, its precession is speed up by $\vec{B}_J^{(i)}$ and if $\vec{S}^{(i)}$ precesses faster than $\vec{B}_J^{(i)}$, its precession is slowed down by $\vec{B}_J^{(i)}$.
In this way, other dephasing effects are  compensated and the dephasing time increases.

To demonstrate this effect, we analyze a minimal toy model comprising two interacting electron spins that are both parallel aligned initially and are strongly coupled to each other.
We start with the simplified Hamiltonian
\begin{align}
	H = J_{1,2} \,\vec{S}^{(1)} \cdot \vec{S}^{(2)} + \vec{b}^{(1)} \vec{S}^{(1)} + \vec{b}^{(2)} \vec{S}^{(2)}
\end{align}
with the dimensionless frozen effective magnetic field
\begin{align}
\vec{b}^{(i)} = \vec{b}_\mr{ext}^{(i)} + \vec{b}_\mr{N}^{(i)} \; .
\end{align}
First, we show that the electron spins do not precess around their local magnetic field $\vec{b}^{(i)}$, but rather around the averaged magnetic field $\vec{b}_+ = ( \vec{b}^{(1)} + \vec{b}^{(2)})/2$.
Second, we demonstrate that the fluctuations of the averaged magnetic field $\vec{b}_+$ 
are smaller than the fluctuations of the individual magnetic fields $ \vec{b}^{(i)}$.
As a result, the dephasing is significantly reduced in presence of a
strong interaction $J_{1,2}$.

In order to highlight the spin precession around the averaged magnetic field, we perform a transformation into the rotating frame of $\vec{b}_+$.
The new Hamiltonian $H'$ in the rotating frame is
\begin{align}
	H' = J_{1,2} \vec{S}'^{(1)} \cdot \vec{S}'^{(2)} + \vec{b}_- (\vec{S}'^{(1)} -  \vec{S}'^{(2)})
\end{align}
with the transformed spins $\vec{S}'^{(i)}$ and the magnetic field difference 
$\vec{b}_- = 
 (\vec{b}^{(1)} - \vec{b}^{(2)})/2$ .
The classical equations of motion in the rotating frame are given by
\begin{align}
\frac{\mr{d}}{\mr{d} t} \vec{S}'^{(1)} &= \left( \vec{b}_- + J_{1,2} \vec{S}'^{(2)} \right) \times \vec{S}'^{(1)}\\
\frac{\mr{d}}{\mr{d} t} \vec{S}'^{(2)} &= \left( - \vec{b}_- + J_{1,2} \vec{S}'^{(1)} \right) \times \vec{S}'^{(2)} \; .
\end{align}
We assume that both spins are almost aligned after the pump pulse 
except for a small deviation {$\vec{\beta}$ $({|\vec{\beta}| \ll 1})$
\begin{align}
\vec{S}'^{(1)} &= \frac{1}{2}\left( \vec{e}_z + \vec{\beta }\right)\\
\vec{S}'^{(2)} &= \frac{1}{2}\left( \vec{e}_z - \vec{\beta }\right) \; .
\end{align}
The initial conditions can be inserted into the equations of motion
\begin{align}
\frac{\mr{d}}{\mr{d} t} \vec{S}'^{(1)}(0)
&= \frac{1}{2} \left( \vec{b}_- \times \vec{e}_z + \vec{b}_- \times \vec{\beta} + J_{1,2} \vec{e}_z \times \vec{\beta} \right) \label{eq:deltazero} \; .
\end{align}
The time derivation $\frac{\mr{d}}{\mr{d} t} \vec{S}'^{(1)}(0)$ in Eq. \eqref{eq:deltazero} is zero for 
\begin{align}
\vec{\beta} = \frac{\vec{b}_-}{J_{1,2}} \;
\end{align}
and therefore, no spin dynamics in the rotating frame occurs.
Note, for $J_{1,2} \gg |\vec{b}_-|$ the deviation $|\vec{\beta}|$ becomes zero, so the spins are parallel aligned.
Thus, strongly coupled parallel aligned spins precess around the averaged magnetic field $\vec{b}_+$
in the limit of an infinitely strong coupling.

\begin{figure}[t]
\begin{center}
\includegraphics[]{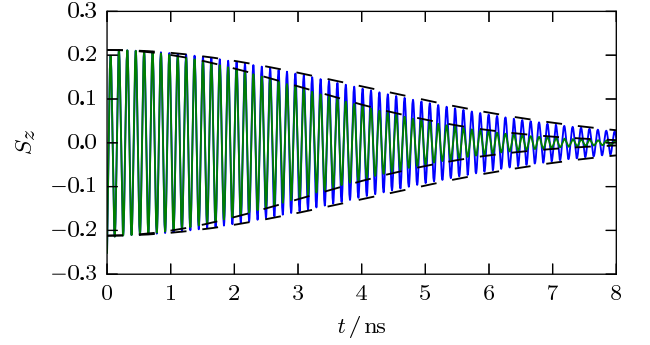}
\caption{
Testing the analytical results of the toy model with a numerical calculation.
Two QDs are excited by an ideal $\pi$-pulse at $t=0$. 
The dephasing time with a strong interaction ($\overline{J}=\SI{200}{ns^{-1}}$, blue curve) is $\sqrt{2}$ times longer than the dephasing time without interaction ($\overline{J}=0$, green curve).
The parameters $T^*_\mr{N}=3\,\mr{ns}$, $g^{(i)}=0.555$ and $B_\mr{ext}=1\,\mr{T}$ are used.}
\label{fig:toy}
\end{center}
\end{figure}

\begin{figure*}
\includegraphics[scale=1]{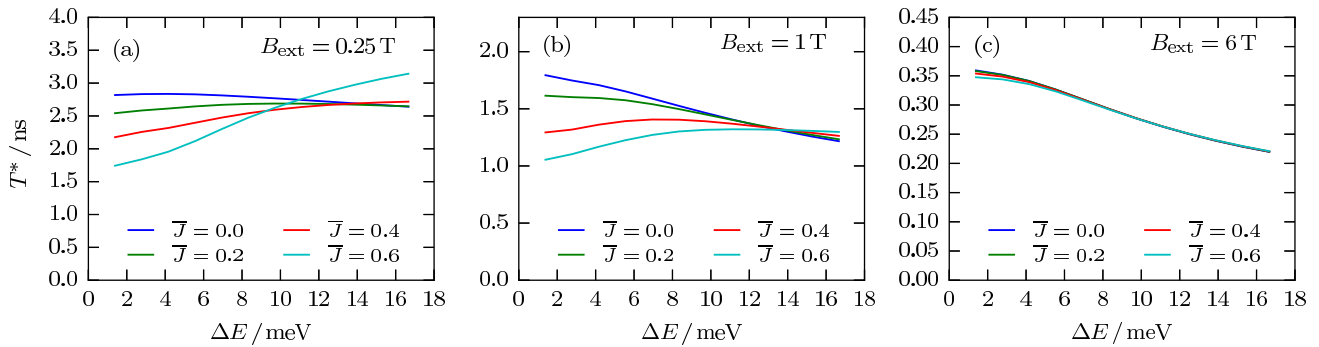} 
\caption{Dephasing time as a function of the spectral width of the laser for three different magnetic fields \SI{0.25}{T} (a), \SI{1}{T} (b) and \SI{6}{T} (c). All three dephasing effects are included in the numerical calculation. There is a good agreement for $\overline{J} = \SI{0.4}{ns^{-1}}$ to the experimental data (c.f. Fig. \ref{fig:fig4}).}
\label{fig:results}
\end{figure*}

As a next step, we consider how the dephasing time is affected
by the noise in $\vec{b}^{(i)}$.
We assume $\vec{b}^{(1)}$ and $\vec{b}^{(2)}$ to have the same fluctuation $\sigma_b=\sigma_{b^{(1)}}=\sigma_{b^{(2)}}$.
To calculate the fluctuations of $\vec{b}_+$ we utilize that $\sigma_{b_+}^2$ and $\sigma_{b^{(i)}}^2$ are cumulants
\begin{align}
	\sigma_{b_+}^2 &= \frac{1}{4} \left( \sigma_{b^{(1)}}^2 + \sigma_{b^{(2)}}^2 \right)
	= \frac{1}{2} \sigma_b^2 \\
	\Rightarrow \sigma_{{b_+}} &= \frac{1}{\sqrt{2}} \sigma_b \; .
\end{align}
The fluctuations $\sigma_{{b_+}}$ of the averaged magnetic field $\vec{b}_+$ are smaller by a factor $\sqrt{2}$ than the single fields  $\vec{b}^{(i)}$.
Thus, the spin dephasing time is reduced due to the electronic spin-spin interaction.
This behavior is demonstrated in Fig.\ \ref{fig:toy}.
Both electron spins are excited by a resonant $\pi$-pulse.
The dephasing time of strongly interacting QDs (blue curve) is longer by a factor 
$\sqrt{2}$ than the dephasing time of uncoupled QDs (green curve).

In conclusion, the effect of the electronic spin-spin interaction on the dephasing time depends on whether the electron spins are polarized or unpolarized.
Unpolarized electron spins create a random magnetic field $\vec{B}_J^{(i)}$ that increases the dephasing while polarized electron spins create a directed magnetic field $\vec{B}_J^{(i)}$ that compensates other dephasing effects.
Hence, the electronic spin-spin interaction generates a faster dephasing for spectrally narrow laser pulses and a slower dephasing for spectrally broad laser pulses.
Note this behavior is contrary to the dephasing due to electron $g$-factor dispersion, where the dephasing time is shorter for spectrally broad laser pulses and longer for spectrally narrow laser pulses.

\subsection{Combining all dephasing effects}
\label{sec:results}

In the previous sections, we identified three contributions to the electron spin dephasing with different signatures.
The nuclear spin fluctuations lead to a dephasing time, that is independent of the laser band width and the external magnetic field.
The electron $g$-factor dispersion yields a faster dephasing for spectrally broad laser pulses and strong magnetic fields.
For the electronic spin-spin interaction, we extracted two regimes that imply a faster dephasing for spectrally narrow pulses and slower dephasing for spectrally broad pulses.
After analyzing the components of the electron spin dephasing separately, we continue by combining the three dephasing effects and match the results to the experimental measurements.
For this purpose, we use the parameters that were already discussed before.
Fig.\ \ref{fig:results} summarizes the combined dephasing time extracted from the full numerical
simulation of the coupled equations.
The three panels present the total dephasing time $T^*$ as function of  
the spectral laser width $\Delta E$ for three different external magnetic fields, $B_\mr{ext} =0.25, 1, \SI{6}{T}$.
Various colors show $T^*$ for different averaged Heisenberg coupling constants $\overline{J}$.
Since we always adjusted the pulse area in the calculations such that a resonant pulse corresponds to 
a $\pi$-pulse, the simulation can be directly compared to the experimental data 
in Fig.\ \ref{fig:fig4} with adjusted power (red curves). 

Without the electronic spin-spin interaction ($\overline{J}=0$, blue curves in Fig.\ \ref{fig:results}) the dephasing time is shorter for spectrally broader pulses.
This behavior is observed for all three magnetic fields.
Furthermore, the dephasing time is reduced with increasing the magnetic field.
This effect is caused by the dephasing due to the electron $g$-factor dispersion
as discussed in detail in Sec.\ \ref{sec:Tg}.

Without the electronic spin-spin interaction, Fig.\ \ref{fig:results} reveals substantial differences
to the experimental data.
Only the results at $B_\mr{ext} = \SI{6}{T}$ match the experiment: The physics is dominated by the Zeeman energy at large magnetic fields.
There is a pronounced negative slope in the numerical results 
as function of $\Delta E$ for all three magnetic fields while in the experiment the dephasing time only exhibits a negative slope at \SI{6}{T}.
For the two smaller magnetic fields, the experimental values of $T^*$ are nearly independent of the laser band width.

With a finite electronic spin-spin interaction, the behavior of the dephasing time changes.
The dephasing time decreases for narrow pulses and increases for broad pulses as analyzed in Sec.\ \ref{sec:TJ}.
While the electronic spin-spin interaction has a strong effect for smaller magnetic fields, the effect is less pronounced at higher magnetic fields where the dephasing contribution due to the electron $g$-factor dispersion dominates.
Especially for an external magnetic field of $\SI{6}{T}$, the effect of the spin-spin interaction almost vanishes.
Only for very narrow laser pulses a small deviation of the curves in Fig.\ \ref{fig:results} (c) is visible.

\begin{figure}
\includegraphics[]{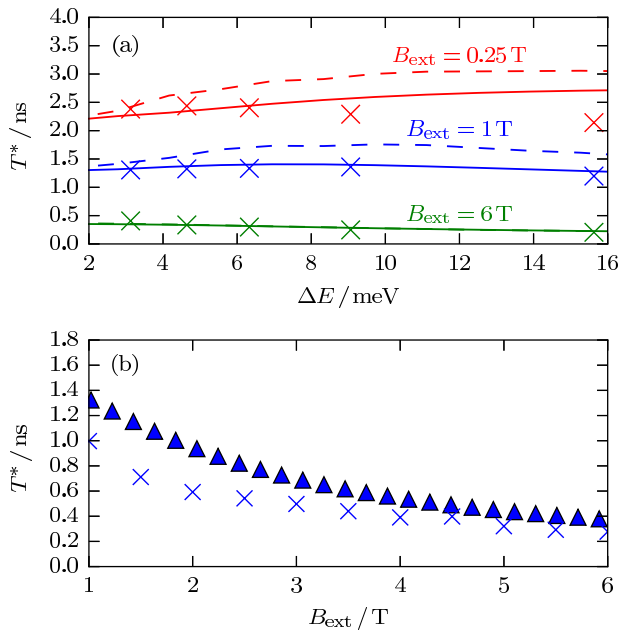}
\caption{Comparison between the numerical results for $\overline{J} = \SI{0.4}{ns^{-1}}$ after a single laser pulse (solid lines) and the experimental measurements from Sec. \ref{sec:experiment} (x markers).
(a) Dephasing time $T^*$ as function of the spectral laser width $\Delta E$.
Results for various external magnetic fields are depicted in different colors.
In addition to the numerical results after a single pulse, we added the calculations after 100 pulses as dashed lines.
(b) Dependence of $T^*$ on the external magnetic field $B_\mr{ext}$
}

\label{fig:combined-data}
\end{figure}

\begin{figure}[t]
\includegraphics[]{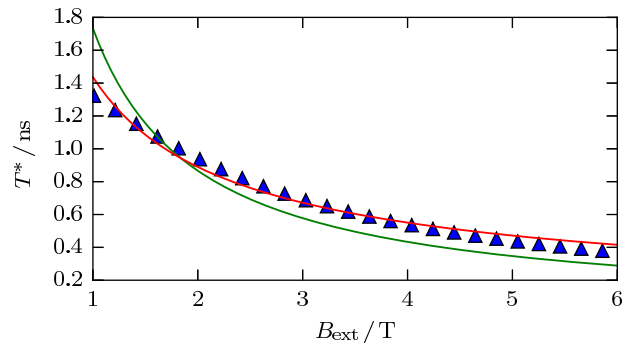}
\caption{
Dephasing time as a function of the magnetic field $B_\mr{ext}$. All three dephasing effects are included in the numerical calculation with $\overline{J} = \SI{0.4}{ns^{-1}}$ and $\Delta E = \SI{1.5}{meV}$. The data points of the numerical calculation (blue triangles) are compared to a dependency $\propto \sfrac{1}{B_\mr{ext}}$ (green curve) and dependency $\propto \sfrac{1}{B_\mr{ext}^{\alpha}}$ with $\alpha = 0.7$ (red curve). }
\label{fig:resultsB}
\end{figure}

Whether the slope of $T^*$ as function of $\Delta E$ is positive or negative is a competition between the electron $g$-factor dispersion (negative slope) and the electronic spin-spin interaction (positive slope). 
Thus, the slope depends on the strength of the external magnetic field $B_\mr{ext}$ and the interaction strength $\overline{J}$.
The best agreement between the experimental data and the numerical simulations
is obtained for $\overline{J}=\SI{0.4}{ns^{-1}}$.
In Fig.\ \ref{fig:combined-data} (a), we combined 
the experimental data shown in Fig.\ \ref{fig:fig4} with the 
calculated values of $T^*$
for  $\overline{J}=\SI{0.4}{ns^{-1}}$  as depicted in Fig.\ 
\ref{fig:results}.
The value $\overline{J}$ reproducing the experimental features best is almost a factor 4 smaller than the value $J\approx \SI{1}{\mu eV} \approx \SI{1.5}{ns^{-1}}$ reported in Ref.\ \cite{Spatzek2011}. 
This smaller value of $\overline{J}$ is a consequence of modeling the ensemble by a cluster of ten QDs instead of a pair only.

The magnetic field dependency of the dephasing time $T^*$ at a fixed spectral width of $\Delta E = \SI{1.5}{meV}$ is depicted in Fig.\ \ref{fig:combined-data} (b).
The numerical calculations with $\overline{J}=\SI{0.4}{ns^{-1}}$ (triangles) as well as the experimental measurements taken from Fig.\ \ref{fig:fig5}(x markers) display a decrease of the dephasing time with increasing external magnetic field.
For our choice of parameters, the numerical values of $T^*$ are slightly bigger than in experiment, but show the same curvature.
In Fig.\ \ref{fig:resultsB}, we supplement a power fit $T^* \propto B_\mr{ext}^{-\alpha}$ to the numerical data.
Like in the experimental results, the best agreement was achieved for $\alpha = 0.7$.
Deviations from the power law with $\alpha=1$, that is predicted by the electron $g$-factor dispersion in Eq. \eqref{eq:Tg}, can be attributed to the importance of the hyperfine interaction and the electronic spin-spin interaction at lower magnetic fields. 

As a next aspect, we analyze the effect of periodic pulse sequences on the dephasing time in our numerical calculations.
The periodicity of the laser pulses imprints on the electron spin dynamics.
As a consequence, the spin system synchronizes with the repetition rate of the optical excitation.
This process happens in two steps \cite{Beugeling2016}:
First, the electron spin synchronizes while the nuclear spins remain unaffected.
To reach this purely electronic steady-state, only a few laser pulses are necessary.
For resonant $\pi$-pulses, a saturation is achieved in less than 10 pulses \cite{Kleinjohann2018}.
In contrast, up to 100 pulses are required to reach the electronic steady-state for detuned QDs, since the pulses are less efficient \cite{Evers2018}.
Second, on a much longer time scale of several thousands or millions of pulses, the nuclear spins align in such a way that the electron spin performs an integer or a half-integer number of revolutions during a pulse interval.
This is called nuclei-induced frequency focusing \cite{Beugeling2016,Jaeschke2017,Kleinjohann2018}.
Due to CPU-time limitations, we restrict ourselves to the analysis of the purely electronic steady-state and leave the effect of the nuclei-induced frequency focusing onto the dephasing time to future investigations.

In Fig.\ \ref{fig:combined-data}(a), we added the dephasing time after 100 pulses as dashed lines.
To save computational effort, we used homogeneous coupling constants ($A_k^{(i)} = A$).
For a periodically pulsed system, the dephasing time increases slightly.
We attribute this increase to the following effect:
The periodic excitation increases the spin polarization of the electron spins, that are affected by the laser pulses.
In this way, the strength of the aligned component of the effective field $\vec{B}_J^{(i)}$ is increased and the dephasing time is extended (see Sec.\ \ref{sec:TJ}).
A better quantitative agreement could be achieved by slightly adapted choices of the parameters, such as $T_\mr{N}^*$.

\section{Conclusion}

\label{sec:conclusion}

We present the dependency of the electron spin dephasing time $T^*$ measured on an ensemble of QDs on the external magnetic field $B_{\rm ext}$ as well as on the spectral width $\Delta E$ of the pump laser.
Although $T^*\propto \Delta E^{-1}$ is expected by the distribution of the electron $g$-factors, the dephasing time is almost independent of $\Delta E$.
Furthermore, the power law $T^* \propto B_{\rm ext}^{-\alpha}$ with a reduced exponent $\alpha\approx  0.7$ can be fitted to the magnetic field dependent data that deviates significantly from $\alpha=1$ predicted from the Zeeman term.

In order to provide an explanation for these observations and reveal the competing mechanisms, the QD ensemble is modeled by a CSM for each singly-charged QD and a static Heisenberg interaction between each pair of electron spins.
Since the microscopic origin is unclear, we included the distance dependency in a simplified exponential distribution of Heisenberg terms $J_{i,j}$ as introduced in Ref.\ \cite{Spatzek2011}.
The electron $g$-factors are Gaussian distributed around a trion excitation energy dependent mean $g$-factor $\overline{g}(\epsilon_\mr{T}^{(i)})$.
We employ the semiclassical approach developed in Ref.\ \cite{Jaeschke2017} combining the classical spin dynamics \cite{Chen2007} with a quantum mechanical description
of the laser pump pulse and extended it to a cluster of coupled QDs to simulate the system of interest.
A key ingredient of this approach is the inclusion of detuning of the laser pulse with respect to the trion excitation energy of each individual QD.
The self-averaging of a large number of configurations is used to reduce the number of
QDs in the cluster to a computationally manageable size by assigning each configuration not only random initial electron and nuclear spins on a Bloch sphere but also random coupling constants in accordance with their distributions.

By benchmarking our approach with the experiment on the optical control of coherent interactions between electron spins in QD ensembles \cite{Spatzek2011}, we have not only validated our simulation but also confirmed that in order to understand the experimental data a fixed and time-independent Heisenberg coupling between electron spins in pairs of QDs is essential to reproduce the experimentally found dependency of $T^*$.

We discuss how each individual effect, the  hyperfine interaction, the electron $g$-factor distribution, the spectral width of the laser and the Heisenberg coupling,
contributes to the dephasing time $T^*$ and derive explicit expressions in certain limits illustrating in detail the physical mechanism.

In the last section, we investigate the combined influence of all the individual contributions, using the estimated parameters from the experiment but vary the mean Heisenberg coupling that is still of unknown microscopic origin.
We found an excellent qualitative and quantitative agreement between the numerical simulations and the experimental data presented at the beginning of the paper when choosing the mean as $\overline{J}=\SI{0.4}{ns^{-1}}$.

In the simulations, we find a partial compensation between two effects at a fixed magnetic field.
On one hand, the dephasing time decreases due to the spread of the electron $g$-factors with increasing $\Delta E$.
On the other hand, $T^*$ increases as a result of the synchronization of an increasing number of QDs in the presences of a static Heisenberg coupling between the QDs.
The total dephasing time $T^*$ appears almost independent of $\Delta E$.

The dephasing time as function of the external magnetic field exhibits a behavior $\propto B_\mr{ext}^{-\alpha}$ ($\alpha < 1$), while a dephasing originating only from the distribution of the electron $g$-factors would predict $\alpha=1$ as can be seen in Eq.\ \eqref{eq:Tg}.
At lower magnetic fields, the importance of the hyperfine coupling as well as the electronic spin-spin coupling increases.
For the parameters used here, the modified power law of the magnetic field dependence yields $\alpha=0.7$ matching the experimental data.

\begin{acknowledgments}
  We are grateful for fruitful discussions on the project with V. V. Belykh, I. Kleinjohann and N. J\"aschke. We acknowledge the supply of the quantum dot samples by D. Reuter and A. D. Wieck (Ruhr-University Bochum). 
We also acknowledge financial support by the Deutsche Forschungsgemeinschaft and the
Russian Foundation of Basic Research through the transregio TRR 160 within the Projects No. A1 and A4.
The authors
gratefully acknowledge the computing time granted by
the John von Neumann Institute for Computing (NIC)
under project HDO09 and provided on the supercomputer JUQUEEN at the J\"ulich Supercomputing Centre.

\end{acknowledgments}


%

\end{document}